\documentclass{elsart}

\usepackage{epsfig}
\usepackage{amsfonts}
\usepackage{amssymb}
\usepackage{amsmath}
\usepackage{bm}

\newcommand{\be}{\begin{equation}}
\newcommand{\ee}{\end{equation}}
\newcommand{\ba}{\begin{array}}
\newcommand{\ea}{\end{array}}

\begin{document}

\begin{frontmatter}
\title{Long Term Economic Relationships from Cointegration Maps}



\author[1]{Renato Vicente}\ead{rvicente@usp.br}\author[1]{, Carlos de B. Pereira}\author[2]{, Vitor B.P. Leite}
\author[3]{ and Nestor Caticha}

\address[1]{GRIFE, Escola de Artes, Ci{\^e}ncias e Humanidades, Universidade de S\~ao Paulo, Parque Ecol\'ogico do Tiet\^e, 03828-020, S\~ao Paulo-SP, Brazil}
\address[2]{Dep. de F{\'\i}sica, IBILCE, Universidade Estadual Paulista, 15054-000 S\~ao Jos\'e do Rio Preto - SP, Brazil}
\address[3]{Dep. de F{\'\i}sica Geral, Instituto de F{\'\i}sica, Universidade de S\~ao Paulo, Caixa Postal 66318, 05315-970 S\~ao Paulo - SP, Brazil}

\begin{abstract}
 We employ the Bayesian framework to define a cointegration measure aimed to represent long term relationships between time series. For visualization
 of these relationships we introduce a dissimilarity matrix and a map based on the Sorting Points Into Neighborhoods (SPIN) technique, which has been previously used to analyze large data sets from DNA arrays. We exemplify the technique
 in three data sets:  US interest rates,  monthly inflation rates and gross domestic product growth rates.
\end{abstract}


\end{frontmatter}

\section{Introduction}
Correlations are a central topic in the study of the collective
properties of complex systems, being of particularly practical
importance when systems of economic interest are concerned
\cite{bouchaud}.
Unlike correlation, the idea of cointegration \cite{engle,hamilton} 
brings in a relationship measure  that is long term in nature being
 somewhat related to the concept of  damage spreading in a pair of
 spin models \cite{damage}.  
However, cointegration  has up to now been rather absent from the
description of physical systems and, in particular, from economic
systems studied from a physical perspective.  
 A set of non-stationary time series cointegrate if
there exists a linear combination of them 
that is mean reverting. Plainly speaking, two appropriately scaled time series
cointegrate if in the long term they either  tend to move together or
as mirror images.

Bayesian methods provide a unifying approach to statistics
\cite{jaynes}. They  help  to establish, from clear first
principles, the methods, assumptions and approximations made in a
particular statistical analysis.  A major issue in the study of
cointegration is the detection of cointegrated sets, a problem that
has been extensively dealt with in the econometrics literature both
from classical \cite{hubrich} and Bayesian \cite{koop} perspectives.

 Dealing with extensive volumes of data is a common trend
in several areas of science. The need to sort, cluster, organize,
categorize, mine or visualize large data  sets  brings a perspective
that unifies distant fields, if not at all in aims, at least in
methods. Cross fertilization may promptly provide candidate
solutions to problems, avoiding the need of rediscovery or worst,
just plain non-discovery. Bioinformatics presents a good example,
where the availability of genome, protein and DNA array data has
prompted the proposal by several groups of new methods. From this
repertoire we borrow a method, SPIN \cite{tsafrir}, previously
developed for automated discovery of cancer associated genes.

Our first goal in this paper is to  devise a cointegration measure
for time series of economic interest that is both physically
meaningful and reasonably simple to compute. Our second goal is, by
employing the SPIN method, to emphasize the importance of  
visual organization and presentation of relationship pictures (or
maps) that emerge when complex systems are analysed. 

This paper is organized as follows. In the next section we derive a
cointegration measure employing Bayesian statistics and briefly
discuss the relation between cointegration and correlation and between
the proposed measure and usual unit-root statistics. In section 3 we
use the SPIN method to  introduce the cointegration heat map  as a
visualization tool. In section 4 we exemplify the proposed technique
in three macroeconomic time series: US interest rates (USIR),
inflation rates (IFR) and gross domestic products (GDP). Conclusions
are presented in section 5.  

\section{Cointegration measure}

A pair of time series ${\bm x}_1$ and ${\bm x}_2$ cointegrates
\cite{hamilton} if there exists a linear combination
\begin{eqnarray}
a_1 x_{1,t} + a_2 x_{2,t} +b &=& \epsilon_t \label{eq:cointegra}
\end{eqnarray}
such that the residues ${\bm \epsilon}$ satisfy the following
stationarity condition:
\begin{eqnarray}
\epsilon_{t+1}&=&\gamma\epsilon_t + \eta_t, \label{eq:stationary}
\end{eqnarray}
where $\langle \eta_t \rangle =0$, $\langle \eta_t^2 \rangle=\sigma^2$
and $\gamma<1$. If $\gamma=1$ the residues are non-stationary and if
$\gamma>1$ the system is unstable. Notice that  $\gamma$ is related to 
a time scale  $\tau=1/(1-\gamma)$ for relaxation of $\epsilon_t$ to its long
term mean.

We also assume a budget
constraint taking the form
\begin{eqnarray}
 a_1^2+a_2^2&=&1.
 \label{eq:budget}
\end{eqnarray}
 Since eq.
\ref{eq:cointegra} is linear, we can impose this constraint by
assuming that  $a_1=\sin(\theta)$ and $a_2=\cos(\theta)$  without
loss of generality. Note that the above system has still more
freedom arising from the following symmetry group:
\begin{eqnarray}
x_i&\rightarrow& x_i'= \alpha x_i+y_i,\nonumber\\
b &\rightarrow& b'=\alpha b -a_1 y_1 -a_2 y_2,\nonumber\\
\sigma &\rightarrow&\sigma'=\alpha \sigma,\nonumber\\
\gamma&\rightarrow&\gamma'=\gamma, \label{eq:symmetries}
\end{eqnarray}
which means that we can change the units in which quantities are
measured and add constants $y_i$ without interfering with the cointegration
property. We, therefore, can partially fix the {\it gauge} so
that $x_i \rightarrow x_i'=x_i - \overline{x_i}$, such that the empiric
time series averages are zero. This forces a choice of $b=0$. 

That cointegration and correlation in time series fluctuations are
quite distinct properties can be easily seen with the aid of an
elementary example. Suppose two time series $x_t$ and $y_t$ that orbit
the same random walk $w_t$ as follows:  
\begin{eqnarray}
x_t&=&w_t+\epsilon^x_t \\
y_t&=&w_t+\epsilon^y_t\nonumber\\
w_t&=&w_{t-1}+\eta_t \nonumber,
\end{eqnarray}  
where $\epsilon^x_t$, $\epsilon^y_t$ and $\eta$ are random i.i.d
shocks with zero mean and variances  $\sigma^2_x$, $\sigma^2_y$ and
$\sigma^2_\eta$, respectively. The correlation $\rho$ between the
first differences $\Delta x=x_t-x_{t-1}$ and $\Delta y=y_t-y_{t-1}$
can be easily computed yielding: 
\begin{eqnarray}
\rho&=&\frac{\langle\Delta x \Delta y \rangle}{\sigma_x\sigma_y}=
\frac{\sigma^2_\eta}{\sigma_x\sigma_y}.
\end{eqnarray}
Clearly $x$ and $y$ strongly cointegrate ($\gamma=0$). However the
choice $\sigma_\eta\ll\sigma_x,\sigma_y$ would  imply that the linear
correlation in their  fluctuations are at the same time  very low.

The main ingredients in a Bayesian approach are three. First we need a
model as given by eqs. \ref{eq:cointegra}, \ref{eq:stationary} and
\ref{eq:budget}. Then a noise model to build the likelihood and
finally the priors. The interesting consequence of a group of
invariance as the one described by eq. \ref{eq:symmetries}  is that
it, together with the budget and stability conditions, constrains \cite{jaynes}
the form of the priors to:
\begin{eqnarray}
p(\gamma)&=&\Theta(\gamma)\Theta(1-\gamma), \label{eq:priorgama}
\end{eqnarray}
where $\Theta(\cdot)$ is the Heaviside step function, and
\begin{eqnarray}
p(\sigma)&\propto& \frac{1}{\sigma^2}. \label{eq:priorsigma}
\end{eqnarray}

With these ingredients we can calculate the posterior probability of
$\gamma$ given the residues as:
\begin{eqnarray}
p(\gamma\mid{\bm \epsilon})&\propto&\int_{\sigma_{\mbox{\tiny min}}}^{\infty}d\sigma\;
p({\bm \epsilon}\mid \gamma,\sigma)p(\sigma)p(\gamma), \label{eq:posterior}
\end{eqnarray}
where $\sigma_{\mbox{\tiny min}}>0$ can be made arbitrarily small
without changing the main results to follow.  

Equations \ref{eq:cointegra} and \ref{eq:stationary} combined give
the following likelihood function for the residues:
\begin{eqnarray}
p(\bm{\epsilon}\mid \gamma,\sigma)&\propto&
\prod_{t=1}^{T-1}\frac{1}{\sigma}\exp\left[-\frac{\left(\epsilon_{t+1}-
\gamma\epsilon_{t}\right)^2}{2\sigma^2}\right].
\label{eq:likelihood}
\end{eqnarray}
Performing the integral in eq. \ref{eq:posterior} yields:
\begin{eqnarray}
p(\gamma\mid{\bm
  \epsilon})\propto\Theta(\gamma)\Theta(1-\gamma)\left[\sum_{t=1}^{T-1}
\left(\epsilon_{t+1}-\gamma\epsilon_{t}\right)^2\right]^{-\frac{T-2}{2}}.
 \label{eq:posterior_integral}  
\end{eqnarray}

\begin{figure}
\includegraphics[angle=0,width=0.5\textwidth]
{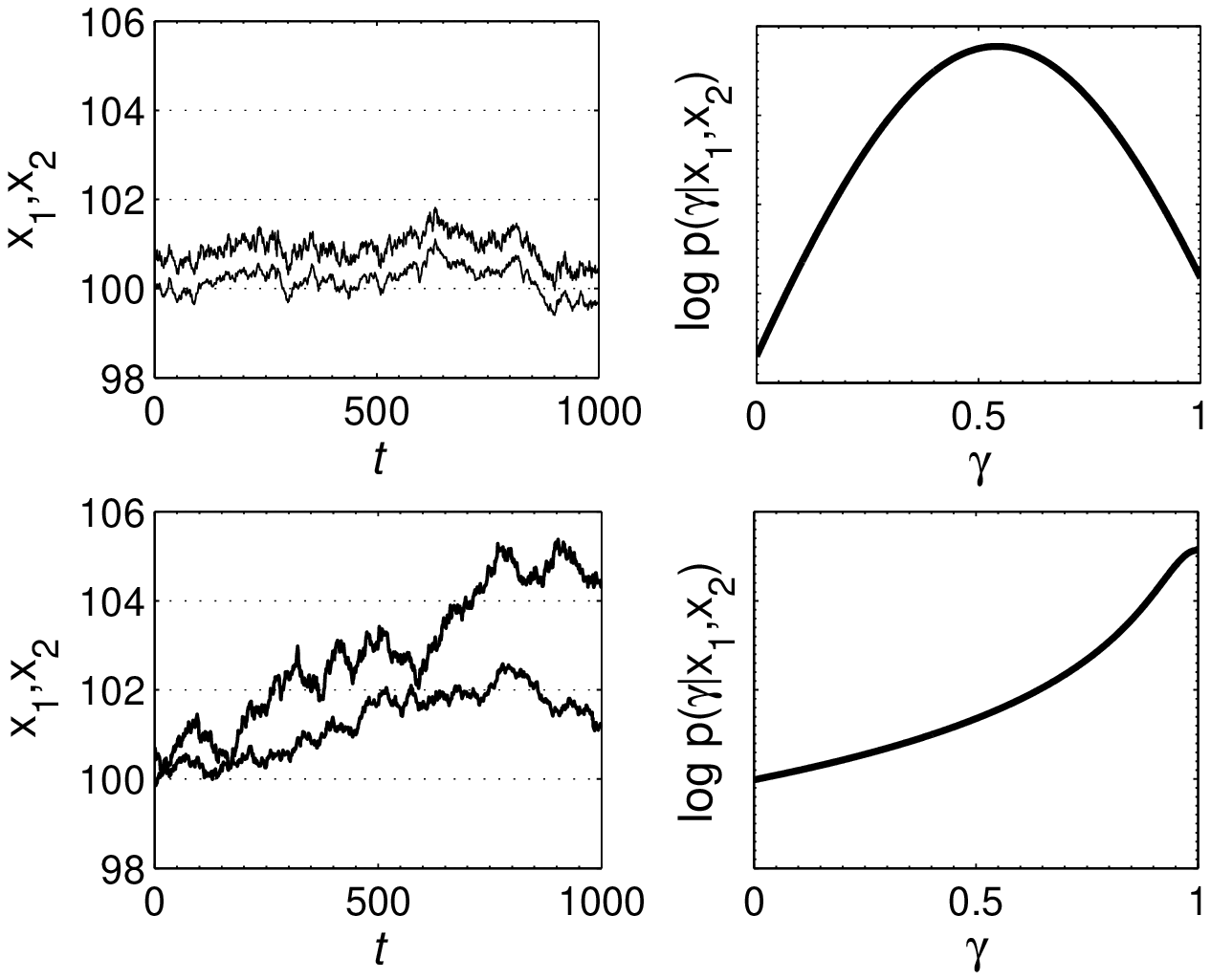}
\includegraphics[angle=0,width=0.5\textwidth]{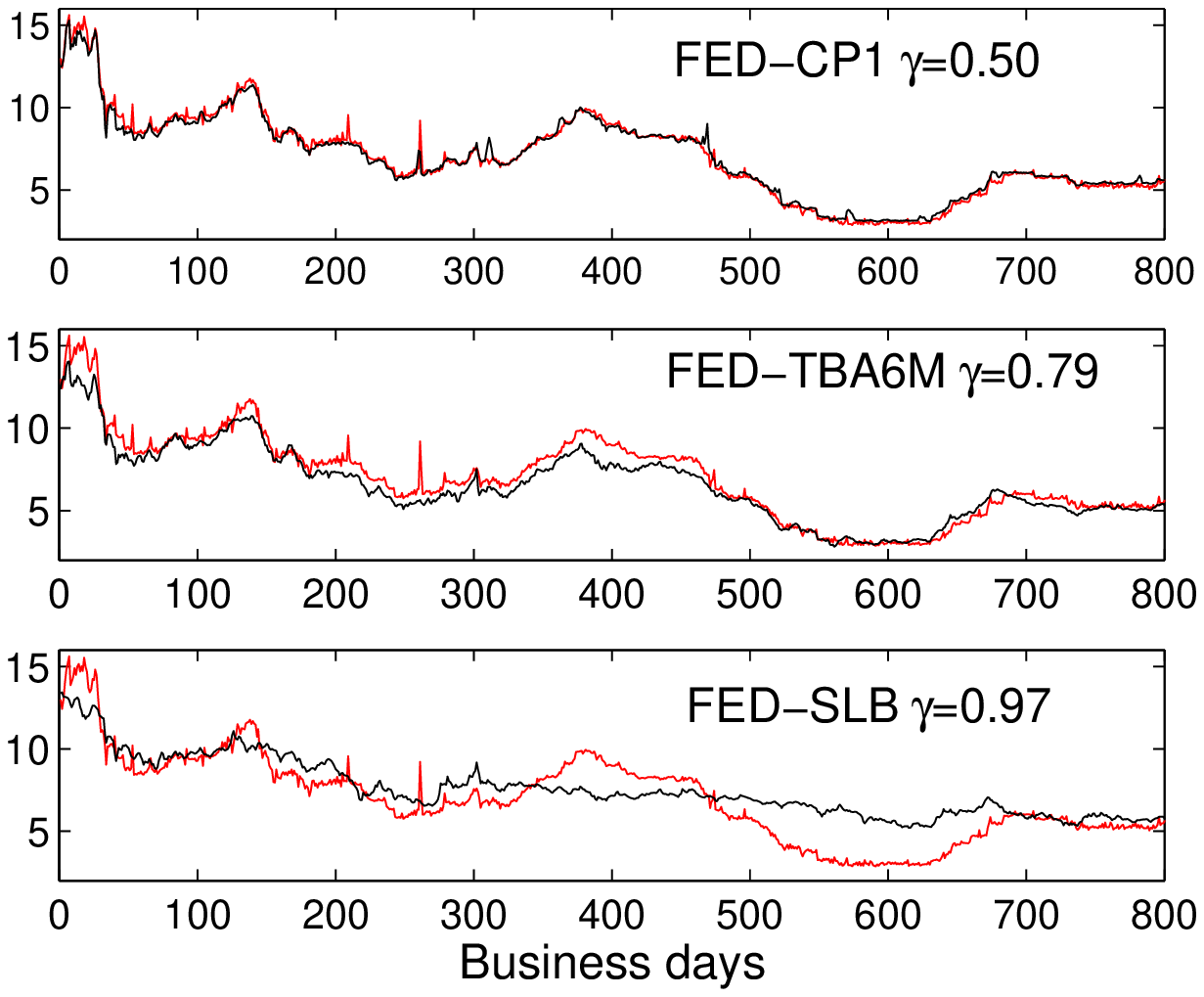}   
\caption{Left: Posterior probabilities for two synthetic pairs ${\bm
x}_1,{\bm x}_2$ of time series. Cointegrating pair,
characterized by the maximum a-{\it posteriori} estimate
$\hat{\gamma}\approx 0.5$ (top). Non-cointegrating pair
characterized by $\hat{\gamma}\approx 1$ (bottom). Right: Examples of
the cointegration measure in US interest rate data.  } \label{fig1} 
\end{figure}

For large $T$, the distribution of residues, given the data, can be
approximated by  

\begin{eqnarray}
p(\bm{\epsilon}\mid {\bm x}^{\prime}_1,{\bm
  x^{\prime}}_2)&\approx&\delta(\epsilon_t-x^{\prime}_{1,t}\sin\hat{\theta} 
+x^{\prime}_{2,t}\cos\hat{\theta})
\end{eqnarray}
with $\hat{\theta}$
estimated by minimizing the variance $\langle\epsilon^2\rangle$ to find:
\begin{eqnarray}
\hat{\theta}&=&
\frac{1}{2}\arctan\left[2\frac{\langle
    x^{\prime}_1x^{\prime}_2\rangle}{\langle {x^{\prime}_1}^2\rangle
    -\langle {x^{\prime}_2}^2\rangle}\right]. 
\label{eq:theta}
\end{eqnarray}

The maximum of the posterior distribution gives an estimate for the
relaxation time  as  
\begin{eqnarray}
\hat{\gamma}&=&\mbox{argmax}
\log p(\gamma\mid {\bm x^{\prime}}_1,{\bm x^{\prime}}_2)\\
		&=&\mbox{argmax}\log\int d\bm{\epsilon}\, 
p(\gamma\mid{\bm \epsilon})p(\bm{\epsilon}
\mid {\bm x}^{\prime}_1,{\bm x^{\prime}}_2)\nonumber.     
\end{eqnarray}
Finally, we define a family of cointegration $\alpha$-measures as
\begin{eqnarray}
d_\alpha({\bm x}_1,{\bm x}_2)\equiv \hat{\gamma}^\alpha.
\end{eqnarray}
These measures are
symmetric, non-negative and agree with the usual augmented
Dickey-Fuller unit-root tests (ADF) \cite{hamilton} in the sense that
lower p-value t-statistics  imply higher degrees of similarity as
measured by the cointegration property (see Table \ref{table}).

The value of $\alpha$ controls the quality of visualizations generated
and has been chosen to be $\alpha=1$ (IFR,GDP) and $\alpha=2$ (USIR)
in the datasets we have analyzed. In Fig.\ref{fig1} (left) we show the
log-posteriors obtained for synthetic time series generated with
$T=1000$ and $\gamma=0.5$ and $\gamma=1.0$. In Fig. \ref{fig1}(right)
we illustrate the cointegration measure with time series from the USIR
dataset. Notice that it can be easily verified by a Taylor expansion
of the logarithm of the posterior density
(eq. \ref{eq:posterior_integral}) around its maximum that the error
bar for the estimate  $\hat{\gamma}$ is  proportional to $T^{-1/2}$.

\section{Cointegration heat map}

\begin{table}
\vspace{0.5cm}
\begin{center}
\begin{tiny}
\begin{tabular}{|c|c|c|} \hline
 \emph{Pair} & \emph{$\hat{\gamma}$} & \emph{ADF t-stat} \\ \hline
SLB-FP3 & 0.99 &   -2.21  \\
CP6-TC1Y& 0.95 &   -4.70  \\ 
CP1-CP6 & 0.92 &   -6.92  \\ 
CP1-CP3 & 0.87 &   -8.72  \\ 
FED-CP6 & 0.77 &   -6.97  \\ 
FED-CP3 & 0.64 &   -8.47  \\
FED-CP1 & 0.50 &   -10.36 \\ \hline
\end{tabular}
\end{tiny}
\caption{ADF test and cointegration measure: Using the USIR dataset as
  an example, it can be seen that the measure estimated are consistent
  with the ADF tests in the sense that more improbable t-statistics
  imply stronger cointegration. The critical value at $1\%$ is
  $t=-3.88$ in this case.\label{table} } 
\end{center}

\end{table}

Given a set of time series we are interested in discovering low
dimensional structures embedded in an appropriate dissimilarity matrix
$D$. The way we define this dissimilarity matrix is conditioned by the
use we intend for the data. In principle, we can define
$D_{jk}=d_\alpha({\bm x}_j,{\bm x}_k)$ meaning that shorter relaxation
times $\tau$  imply more similarity between two time series.
Alternatively, 
inspired by the expression matrices employed in bioinformatics
\cite{tsafrir}, we can define  vectors
$\mathbf{d}_j=(d_{1j},...,d_{Nj})$  representing the cointegration
profile between time series $j$ and each one of the $N$ series
composing the system with an arbitrary but fixed ordering. A
dissimilarity matrix can be then defined along this lines as: 
\begin{equation}
D_{jk}=\sqrt{\sum_{l=1}^N \left(d_{lj}-d_{lk}\right)^2}.
\label{dissimilarity}
\end{equation}  
In  this case two time series are  similar if they interact with each
system component alike. We have observed that the latter choice yields
the same basic structures with clearer and smoother visualization. The
use of primitive measures to build second order dissimilarity matrices
is a basic idea 
behind the spectral clustering techniques \cite{spectral}. However, to
our knowledge, the particular construction described by
eq. \ref{dissimilarity}  has not appeared in the literature to date. 

\begin{figure}
\includegraphics[angle=0,width=0.5\textwidth]{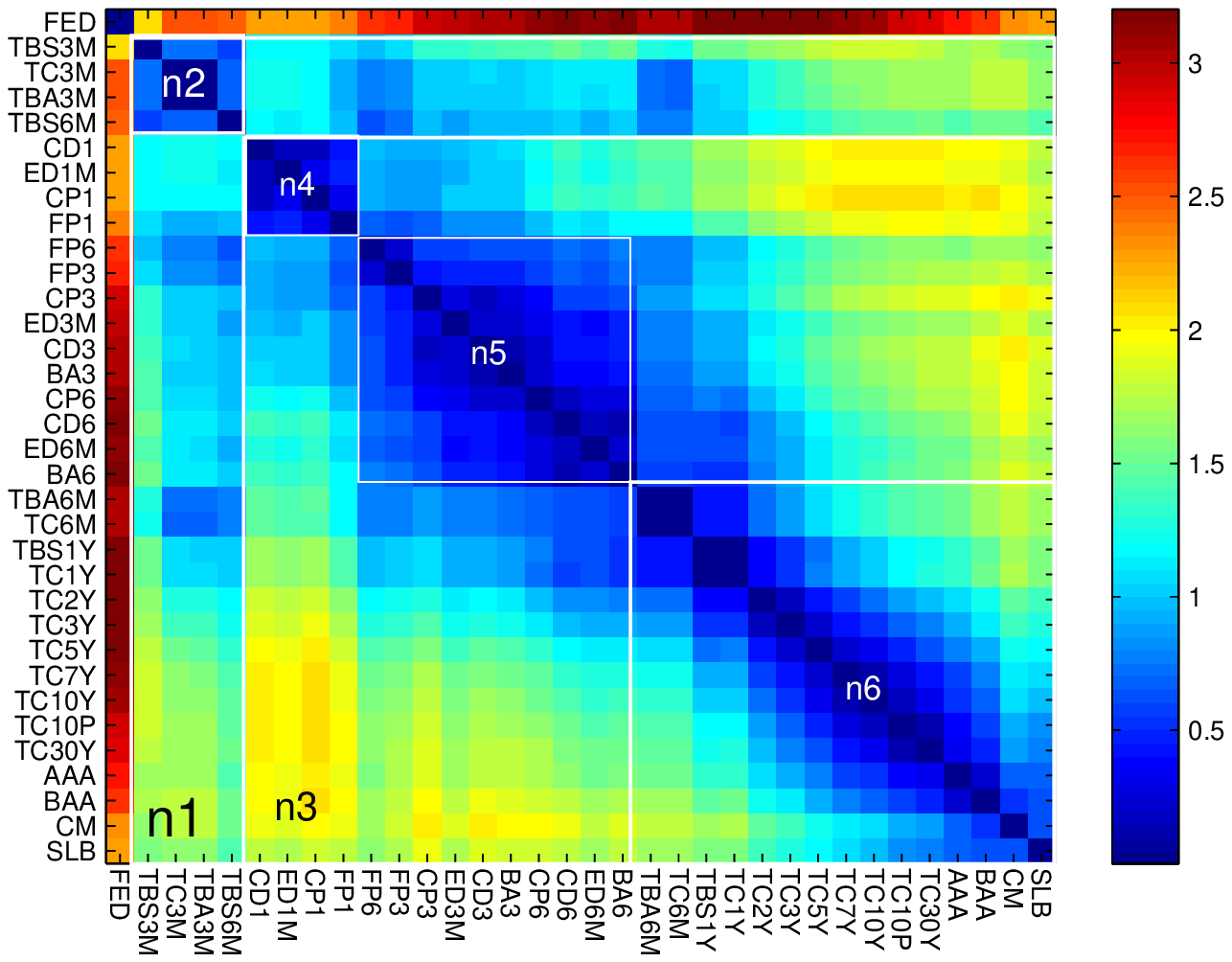}\includegraphics[angle=0,width=0.5\textwidth]{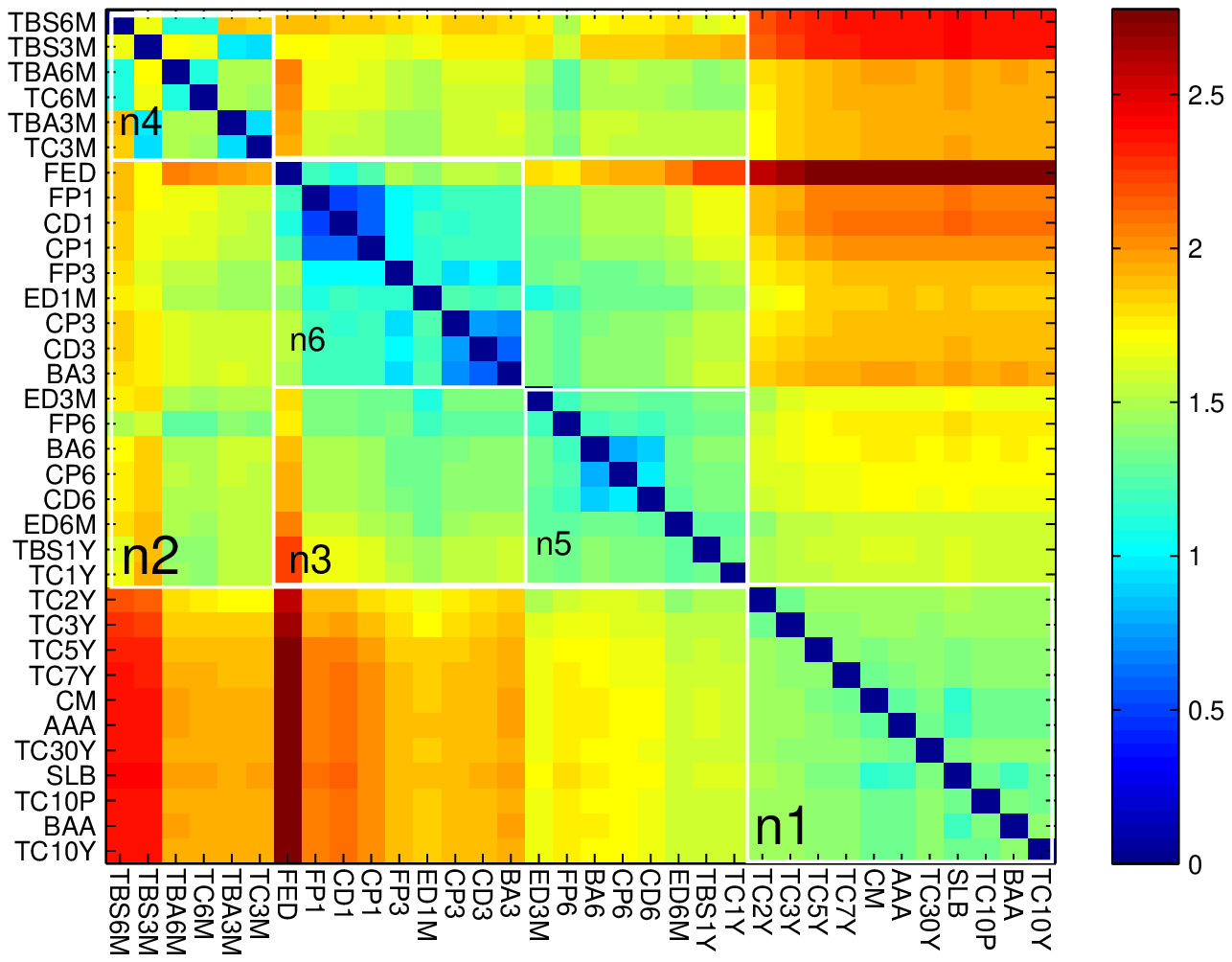}  
\caption{Left: Correlation heat map for the USIR dataset. Rectangles
  represent the classification yielded by the SPC technique. The
  general pattern is comparable with figure 3b on
  \cite{dimatteo_usir}. Right: Cointegration heat map for the USIR
  dataset with SPC classification again represented  by
  rectangles. The general emergent pattern clusters short term
  Treasury instruments around $n4$, financial companies related
  instruments around $n3$ and long term instruments around $n1$.} 
\label{fig2}
\end{figure}

 There are several possible aims behind unsupervised segmentation
based on a dissimilarity  matrix $D$. Categorization from clustering
algorithms has been used for market segmentation
\cite{dimatteo,tumminello}. For example, the 
Superparamagnetic Clustering (SPC) algorithm \cite{SPC,kullmann} has
been particularly useful 
since the number of clusters is not {\it a priori} known and the
scale of resolution of the categories can be tuned by a temperature
like parameter. Sometimes the data might not have a clear discrete
class structure and here the SPIN algorithm provides a difference
with its capability of helping identify low dimensional structures
in a high dimensional space. Without knowing in advance what type of
segmentation will emerge,  the clustering and SPIN  algorithms
should be thought of as complementary. The aim of SPINing a similarity
matrix is to obtain a permutation such that points close in distance
are brought, by the permutation to places in the matrix that are
also close. Since the space of permutations is factorially large
this can easily be seen to be a potentially hard problem. The
permutations are sequentially chosen, for example to  minimize a
cost function that penalizes large distances and puts them far from
the diagonal or alternatively seek permutations that bring pairs
with small distances near to the diagonal. Unless the structure can
be ordered in one dimension, these requirements can lead to
frustration. The class of  cost functions proposed in \cite{tsafrir}
is of the form $\mathcal{F}(P)=${\bf Tr}$(PDP^T W)$, with $P$ being a 
permutation of matrix indices and $W$ a weight matrix which defines
the algorithm.  
For their choices, namely, {\it Side-to-Side (STS)} defined as
$W=XX^T$, with $X_i>X_j$ if $i>j$ and {\it Neighborhood} defined as
$W_{ij}=exp(|i-j|²\sigma²)$, the minimization was shown to be
NP-complete. The way out is to be satisfied with non optimal
solutions that can be obtained in fast times ($\mathcal{O} (n^{2-3})$)
and that turn out to be just as informative. The problem of sorting
into categories is ill posed and therefore there will not be
something like `the answer'. The reduction to an optimization
problem, using either {\it STS} or {\it Neighborhood} leads to a NP-complete
problem. It is fair to expect that any reasonable weight function
will share that characteristic. So we have found that it is adequate
to play around with the algorithms and apply them for different
subsets, try optimizing the whole matrix, then choose a relevant
cointegrating subset, optimize the subset, go up optimize the whole
set, intercalate different algorithms. The result will tend to be
better as measured by the cost function. This heuristics helps
escape from local minima, of course it does not cure the fundamental
problem that there might be frustration in a general sorting
problem. This is not really a problem, good albeit not optimal
solutions are just as informative as a perfect solution would be for
all practical purposes. In the following analysis we have found that
the best visualizations  
were simply achieved by employing the  {\it STS} solution as 
an initial condition to the {\it Neighborhood} variant iterated with a
schedule for 
reducing the scale parameter $\sigma$ in a simulated annealing fashion.

\begin{figure}
\includegraphics[angle=0,width=0.5\textwidth]{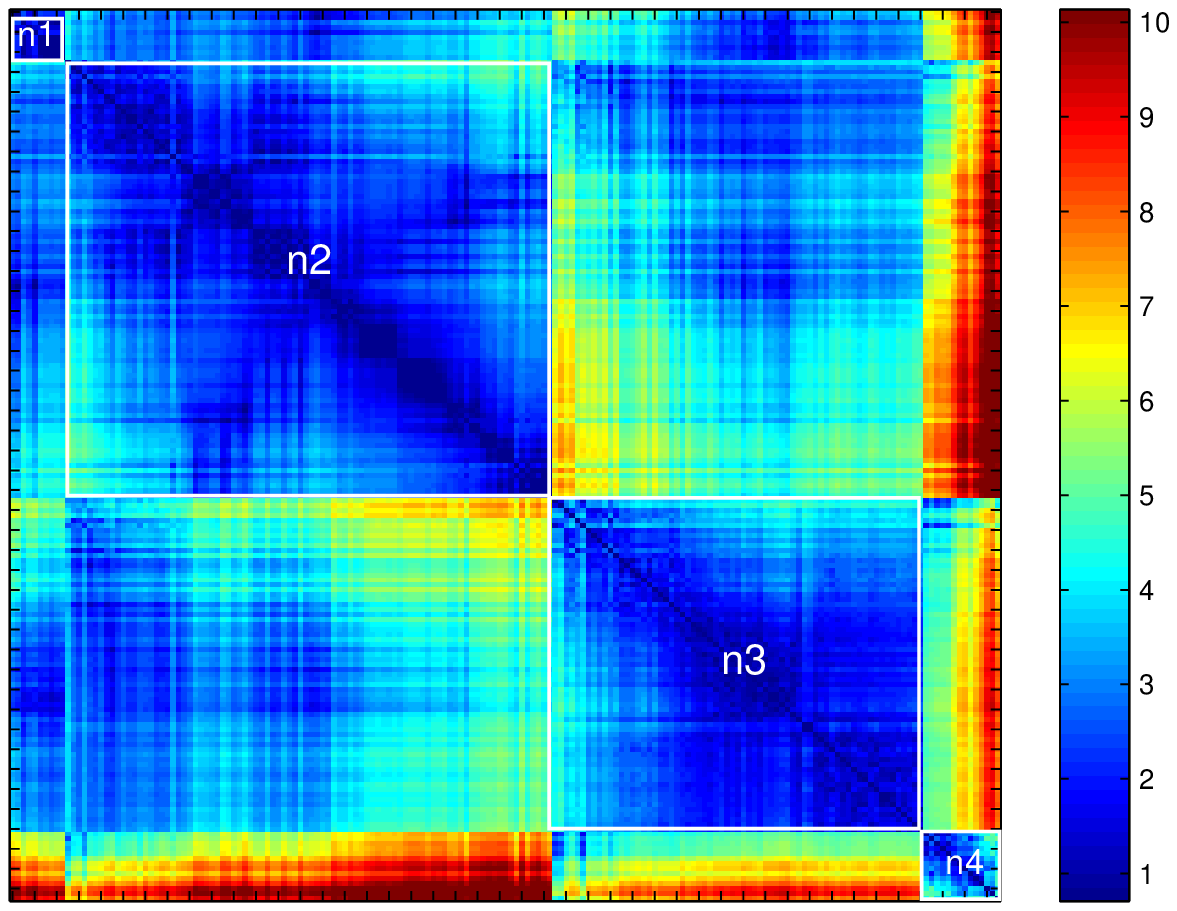}\includegraphics[angle=0,width=0.5\textwidth]{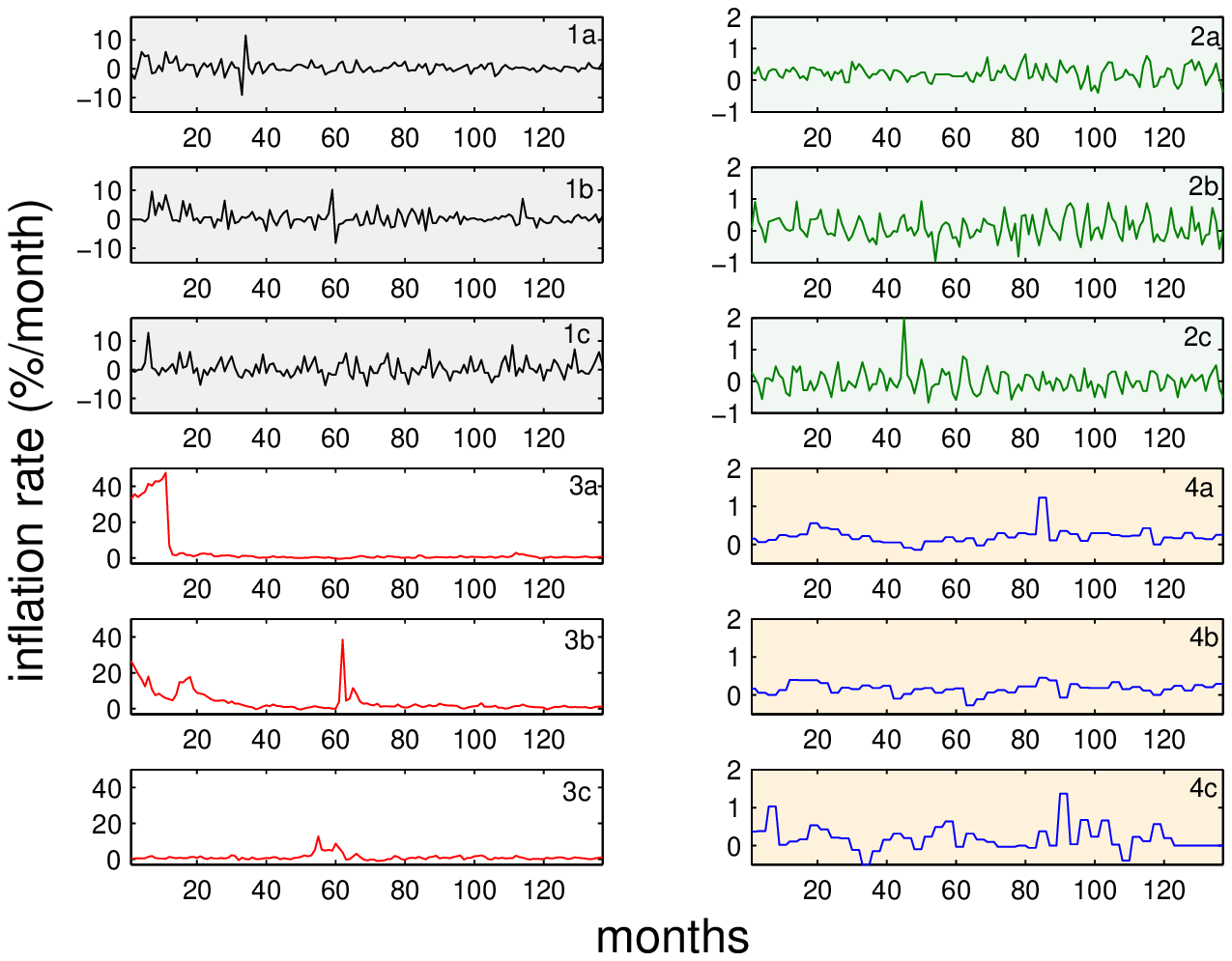}
\caption{Left: SPINed cointegration map for the IFR dataset.  The
  clusters that emerge correspond to economies with highly volatile
  prices (n1), developed economies with stable prices (n2), economies
  with events of hyperinflation in the period observed (n3) and
  economies with highly stable prices (n4). Right: Examples in each
  group. Group n1: Benin (1a), Central African Republic (1b) and Syria
  (1c). Group n2: USA (2a), Sweden (2b), Japan (2c). Group n3: Brazil
  (3a), Russia (3b) and Indonesia (3c). Group n4: Australia (4a), New
  Zealand (4b) and Tuvalu (4c). } 
\label{fig3}
\end{figure}

\section{Application examples}
We exemplify the method by calculating cointegration maps for three
data sets: (USIR) weekly US interest rates for 34  instruments from
January 8, 1982 to August 29, 1997 ($T=817$ weeks) \cite{usinterest};
(IFR)  monthly inflation rates for 179 countries from August, 1993 to
December 2004 ($T=137$ months) \cite{inflation}; (GDP) yearly gross 
domestic product growth rates for 71 countries from 1980 to 2004
($T=25$ years) \cite{gdp}.     

Measurement in soft sciences is itself a challenging activity
\cite{boumans}. Socio-economic systems are self-aware, there are
severe limits to the accuracy of statistical data that can be gathered
and even the definition of several macroeconomic quantities is still
debatable \cite{swanson,morgenstern}.  An exception to these data
quality constraints are the organized financial markets like those of
interest rate instruments in dataset USIR.

In figure \ref{fig2} (left)  we show the SPINed heat map for
correlation coefficients of time series fluctuations.  Pseudocolors
are assigned according to  dissimilarities calculated with
eq. \ref{dissimilarity} by replacing the cointegration measure by
correlation coefficients. In the same figure we show as rectangles
identified by $n1,...,n6$ the hierarchical grouping structure
generated by the SPC technique ($K=7$, see \cite{SPC}).
Characteristic of unsupervised classification techniques is the
reliance of the results on the dissimilarity measure adopted. Despite
the differences, the general patterns revealed in figure \ref{fig2}
compare well with those of figure 3b on \cite{dimatteo_usir}, which
employs a classical agglomerative clustering with a metric distance
based on linear correlation coefficients. Notice that the  SPINed heat
map is capable of showing nuances in the relationship structure that
are absent in the traditional or SPC approaches. For example, the
Treasure bill rates with maturities 3 and 6 months (TBA3M and TBA6M)
and  other instruments of the same maturity, in particular, Treasure
securities at constant maturity (TC3M and TC6M) correlate
alike. However, this sort of information is lost both in the SPC
classification and in \cite{dimatteo_usir} with the former classifying
these instruments accordingly with their maturities and the latter
grouping TBAs in one group and TCs in another.   
  
\begin{figure}
\includegraphics[angle=0,width=0.5\textwidth]{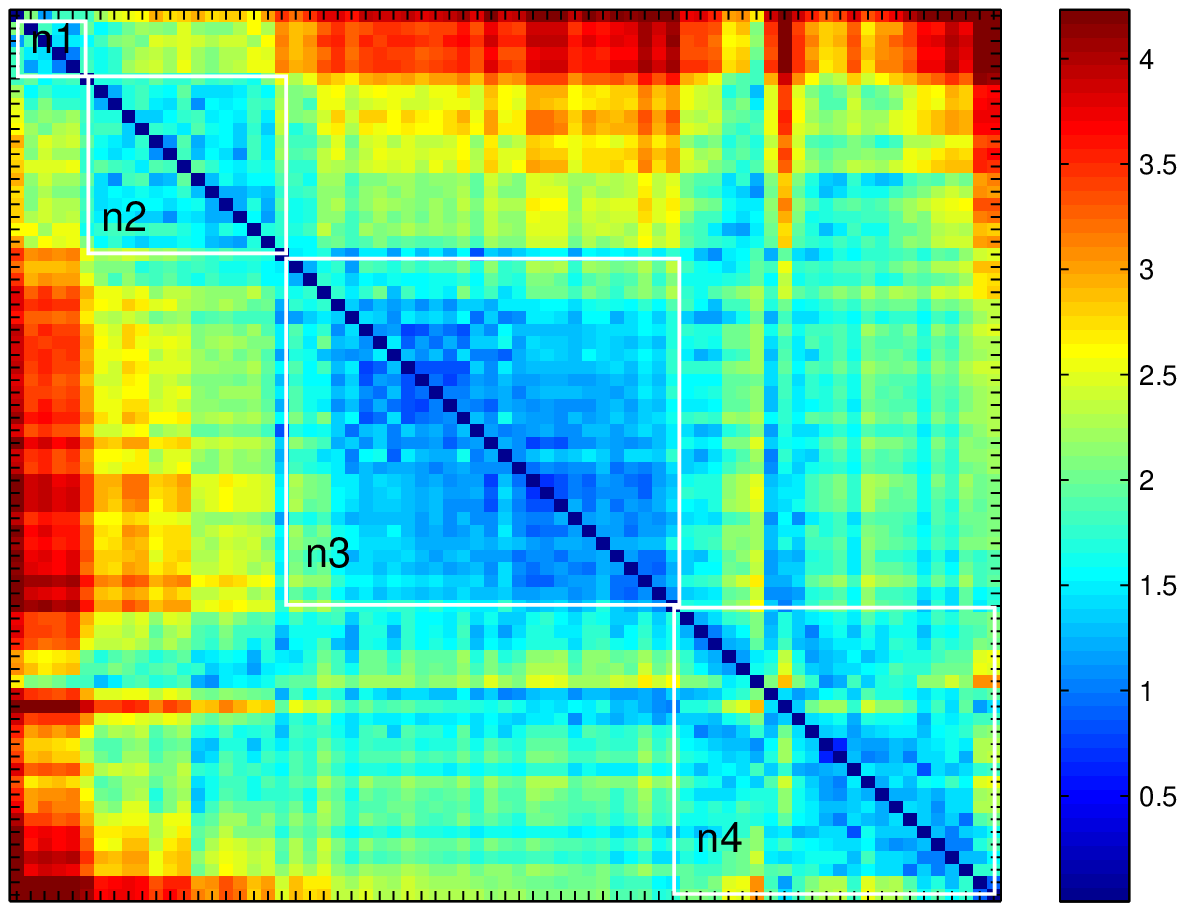}\includegraphics[angle=0,width=0.5\textwidth]{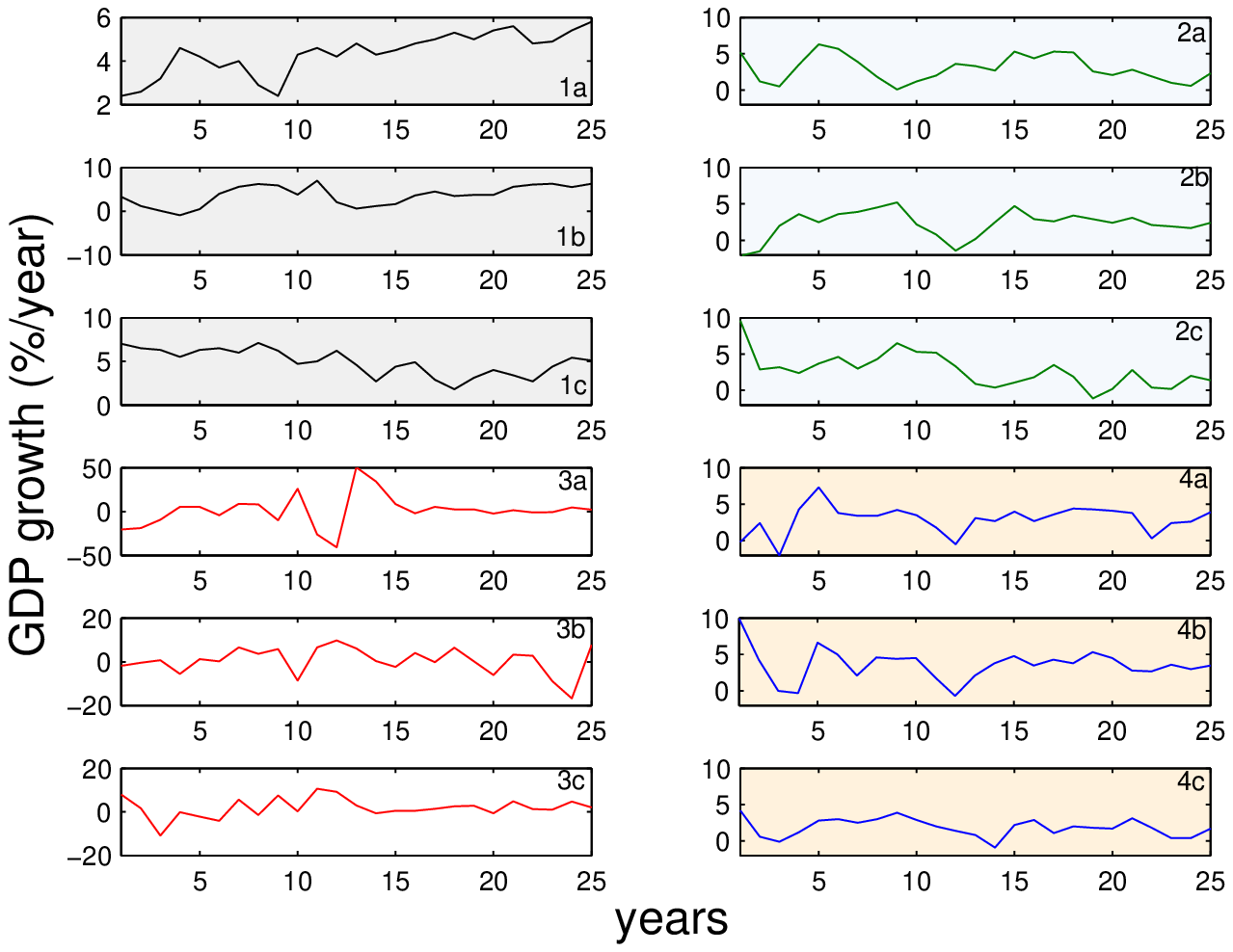}  
\caption{Left: SPINed
cointegration map for the GDP dataset.  The clusters that emerge
correspond to countries with accelerating or decelerating economies
(Group $n1$); developed countries with stable  and  accelerating
economies (Group $n2$); volatile economies including major oil
producers (Group $n3$) and stable economies (Group $n4$). Right:
Examples in each group are. Group $n1$: Bangladesh (1a), Tanzania (1b)
and Pakistan (1c). Groups $n2$: Norway (2a), United Kingdom (2b) and
Japan (2c). Group $n3$: Kuwait (3a), Venezuela (3b) and Saudi Arabia
(3c). Group $n4$: US (4a), Australia (4b) and Italy (4c).} 
\label{fig4}
\end{figure}

Figure \ref{fig2} (right) shows the cointegration map for USIR.
Considering the reliability of the estimates ($\sigma_{\gamma}\approx
0.035$ for USIR), the map produced allows a direct visualization of
relationships through the whole set of time series without imposing
any {\it ad hoc} classification criteria. The classification provided
by SPC ($K=7$) is also shown as rectangles identified by
$n1,...,n6$. As we have already discussed, correlation of the
fluctuations and cointegration are different  relationship
measures. Two time series will pertain to the same cointegration group
if they tend to orbit a common time series. The general pattern groups
short term Treasury bills in $n4$, long term instruments, both treasury
and corporate around $n1$ and finance company related instruments
(interbank eurodollar (EDs), certificates of deposit (CDs) and finance
company papers (FPs)) around $n3$.   

Figure \ref{fig3} (left) shows the SPINed cointegration map for
monthly inflation data (IFR). Even though the estimates are less
reliable in this case ($\sigma_{\gamma}\approx 0.085$) it is possible
to identify groups by inspecting their mutual relationships
represented by the color map. Figure \ref{fig3} (right)  allows a
direct interpretations of segmentation provided. Group $n1$ consists
of countries that exhibit  volatile inflation profiles with both high
inflation and high deflation periods (Benin (1a), Central African
Republic (1b) and Syria (1c)). Group $n2$ is mainly composed by
advanced economies with stable inflation patterns (USA (2a), Sweden
(2b) and Japan (2c)) and countries that are very closely related to
them (e.g. Martinique, Singapore and Bahamas). Group $n3$ consists of
countries that have experienced hyperinflation in the period observed
(Brazil (3a), Russia (3b) and Indonesia (3c)). Group $n4$ contains
countries with very stable and low inflation profiles, among them
New Zealand (4b), that has adopted inflation targeting as early as
1988, Australia (4a), that also has adopted inflation targeting in
1993  and Tuvalu (4c) that adopts the Australian dollar as
currency. However, apart from Australia and New Zealand, all the other
countries that have adopted inflation targeting before the period
observed (1993-2004) (Canada, Finland, Korea, Sweden and United
Kingdom)  have been classified in the Group $n2$.

The cointegration map for GDP data (Fig. \ref{fig4} (left)) must be
dealt with care as  this data set is smaller ($T=25$) and, therefore,
statistically less reliable than the previous two sets
($\sigma_{\gamma}\approx 0.2$). To minimize  interpretation problems
due to GDP measurement issues we have selected from the IMF database
$71$ countries that have had  market economies in the period observed
(1980-2004).  As a criterion to classify different groups, we have
looked at general interaction patterns compatible with the limited
reliability of the estimates. The SPINned matrix shows that there are
four distinguishable classes,  but that their boundaries are not
sharp. This illustrates again the  difference between 
SPIN and traditional clustering techniques. For the latter either
sharp boundaries (e.g. for hierarchical and K-means techniques) or
some sort of low dimensional structure (e.g. fuzzy clustering) must be
imposed even when there are none \cite{hastie}. We, therefore, have
{\it defined} Group $n1$ as being composed by countries that interact
with countries in Group $n2$. Group $n2$ consists of countries that
interact  with Group $n4$ and less strongly with Group $n3$. Group
$n3$ is characterized by countries that do not interact with Group
$n1$, interact strongly with Group $n4$ and less strongly with Group
$n2$. Finally, Group $n4$ interacts with Groups $n2$ and $n3$ but not
with Group $n1$.  This procedure results  in accelerating 
(Fig. \ref{fig4} (right) Bangladesh (1a) and Tanzania (1b) ) or
decelerating economies (Pakistan (1c)) in Group $n1$; low volatility
economies both stable (Norway (2a) and United Kingdom (2b)) and
decelerating (Japan (2c)) in Group $n2$; Group $n3$ concentrates
highly volatile unstable economies including developing countries and
all major oil producers (Kuwait (3a), Venezuela (3b) and Saudi Arabia
(3c)); stable economies in Group $n4$ (US (4a), Australia (4b) and
Italy (4c)). Notice that the difference between Groups $n4$ and $n2$
is their relation with Group $n1$, to say, the presence of some
countries with accelerating or decelerating growth rates in $n2$.

\section{Conclusion}
In this paper we have developed a simple  measure for long term
pairwise relationships in sets of time series  by introducing a
Bayesian estimate for  a cointegration distance. 
For visualization of the relationships, with a minimum introduction of
{\it ad hoc} structures, we have borrowed from the repertoire of
Bioinformatics the SPIN ordering technique to produce {\it
  cointegration heat maps}. We have exemplified the technique  in three     
sets of time series of financial and economic interest and have been
capable of identifying low-dimensional  structures of economic sense
emerging from the procedure.

Our aim in this work has been the development of tools that may be
useful for discovering  collective long term structures in economic
time series. We think  that a thorough understanding of the economic
phenomena behind the observed patterns depends on our capability of
describing the system interactions in some detail, what is out of the
scope of the present work. Considering that socio-economic systems
belong to a class of complex systems with unreliably known
interactions and dynamics we regard pattern recognition tasks as
hereby described a first important step towards a deeper quantitative
understanding of such systems.

\ack
A previous version of this work has been presented at APFA5 in Torino.
We wish to thank Eytan Domany and his collaborators for valuable
comments on the manuscript and  
  for gently providing a shareware license of the  Analyzer and Sorter (SPIN)
package we have employed to build SPINed cointegration heat maps.
RV would like to thank the hospitality and financial support of the
\'Ecole de Physique Les Houches where part of this work has been completed.
VBPL has been funded by {\bf FAPESP} under  research grant 05/58474-1.
A detailed description of the clusters obtained as well as the scripts
used for computing the  
cointegration measure will be available at the webpage of one of the
authors (RV).

%

\end{document}